\documentclass[onecolumn,pra,a4paper,showpacs,floatfix,preprintnumbers,amsmath,amssymb]{revtex4}

\usepackage{graphicx}% Include figure files
\usepackage{dcolumn}% Align table columns on decimal point
\usepackage{bm}% bold math

\newcommand{\ud}{\mathrm{d}}

\newcommand{\be}{\begin{equation}}
\newcommand{\ee}{\end{equation}}

\newcommand{\BI}{\mathrm{I}}

\begin{document}

\title{Sub Shot-Noise Phase Sensitivity with a Bose-Einstein Condensate Mach-Zehnder Interferometer}% Force line breaks with \\

\author{L. Pezz\'e$^{1,2}$, L.A. Collins$^1$ 
A. Smerzi$^{1,2}$, G.P. Berman$^1$, and  A.R. Bishop$^1$}
\affiliation{1) Theoretical Division, Los Alamos National Laboratory,
Los Alamos, New Mexico 87545, USA\\
2) Istituto Nazionale per la Fisica della Materia BEC-CRS\\
and Dipartimento di Fisica, Universit\`a di Trento, I-38050 Povo, Italy}

\date{\today}% It is always \today, today, but any date may be explicitly specified

\begin{abstract} 
Bose Einstein Condensates, with their coherence  properties, have
attracted wide interest for their possible application to ultra precise
interferometry and ultra weak force sensors.  Since condensates, unlike
photons, are interacting, they may permit the realization of specific
quantum states needed as input of an interferometer to approach the
Heisenberg limit, the supposed lower bound to precision phase
measurements.  To this end, we study the sensitivity to external weak
perturbations of a representative matter-wave Mach-Zehnder
interferometer whose input are two Bose-Einstein condensates created by
splitting a single condensate in two parts.  The interferometric phase
sensitivity depends on the specific quantum state created with the two
condensates, and, therefore, on the time scale of the splitting
process. We identify three different regimes, characterized by a phase
sensitivity $\Delta \theta$ scaling with the total number of condensate
particles $N$ as i) the standard quantum limit $\Delta \theta \sim
1/N^{1/2}$, ii) the sub shot-noise $\Delta \theta \sim 1/N^{3/4}$ and
the iii) the Heisenberg limit $\Delta \theta \sim 1/N$. However, in a
realistic dynamical BEC splitting, the $1/N$ limit requires a long
adiabaticity time scale, which is hardly reachable experimentally. On
the other hand, the sub shot-noise sensitivity $\Delta \theta \sim
1/N^{3/4}$ can be reached in a realistic experimental setting.  We also
show that the $1/N^{3/4}$ scaling  is a rigorous upper bound in the
limit $N \to \infty$, while keeping constant all different parameters
of the bosonic Mach-Zehnder interferometer.
 
\end{abstract}

\maketitle

\section{introduction}
In the last few years theoretical and experimental efforts have been devoted to
the realization of an ultraprecise quantum interferometer (for a review see \cite{Giovanetti_2004}).
High resolution phase measurements find applications in the detection of ultraweak forces, 
as gravitational waves \cite{Caves_1981}, and inertial forces \cite{McGuirk_2001}.
The Mach-Zehnder (MZ) interferometer is a prototypical apparatus employing both optical and matter waves. 
In the optical MZ, the phase sensitivity depends on the intensity
of the laser field, which corresponds to a limit on the average number $N$ of photons. 
Typically, the sensitivity is bounded by the shot noise limit $1/\sqrt{N}$, which can be reached with 
classical states (for example vacuum plus coherent state) as input of the interferometer. This limit, 
however, is not fundamental, and it can be surpassed with
the use of nonclassical states exploiting, in this way, quantum correlations.   
The insuperable lower bound to precision phase measurements is believed to be
given by the Heisenberg limit $1/N$ \cite{Ou_1997}.
Its achievement has been theoretically demonstrated in a large body of literature with a 
variety of states and optimal performances 
\cite{Yurke_1986, Boundurant_1984, Holland_1993,Hillery_1993, Bollinger_1996, Barry_2000}. 
Bose Einstein Condensates (BEC), with their coherence  properties, 
have attracted a wide interest for their possible application in ultraprecise interferometry \cite{Bouyer_1997}, and 
ultraweak force sensors \cite{Harber_2005, Antezza_2004,ingu}.
Condensates, unlike photons, are interacting, and this property seems very promising
for the realization of quantum states to use as input of an interferometer to approach the Heisenberg limit.
For example, the recent experimental creation of Fock states \cite{Greiner_2002} and of very stable double-well 
traps \cite{Shin_2004, Oberthaler_2005} bodes well for the future of matter-wave interferometry.

In this paper, we analyze a BEC Mach-Zehnder interferometer. 
The initial state configuration is prepared by trapping a condensate in a double-well potential with an 
interwell barrier large enough to create the two independent condensates that feed the interferometer.
The height of the potential barrier is decreased
instantaneously, and a tunneling between the two condensates is allowed for a time $t_{\pi/2}=\pi/2\Omega$, where 
$\Omega$ is the condensate tunneling rate between the wells. 
The barrier is then increased again in order to have a negligible tunneling rate. During this time, the interaction of 
a weak force with the condensates will shift their relative phase by an amount proportional to the energy gradient
induced by the external field and the time of exposure.
After a second $\pi/2$ pulse, the relative number of particles is measured, and information on the 
phase shift is recorded.
Different measurement schemes have been proposed, based on a positive operator value measurement \cite{Sanders_1995}, 
on a parity measurement \cite{Campos_2003}, on a relative number fluctuation measurement \cite{Kim_1997, Bouyer_1997}, 
and on a collapse and revival detection \cite{Dunningham_2004}.
We study in detail the splitting processes by analyzing the produced quantum states, and 
giving a prediction on the sensitivity of the interferometer by using an error-propagation formula.
We assume, for simplicity, lossless devices (beam splitters). 
However, losses of atoms can degrade the sensitivity back to the shot noise limit \cite{Huelga_1997}. 
With this setup we expect an improvement of phase sensitivity, reachable with classical 
states, toward the quantum Heisenberg limit.
In fact, as analyzed by J\"a\"askel\"ainen, \emph{et al.} \cite{Meystre_2004}, 
for a complete adiabatic split, a repulsive-interaction condensate will end
in a Fock state $|\psi\rangle=|N/2\rangle|N/2\rangle$. With this state as input, the MZ
phase sensitivity is expected to scale at the Heisenberg limit \cite{Holland_1993}.
In this paper, we point out the existence of three different regimes, which, in analogy with three
corresponding regimes existing in the dynamical Josephson effect \cite{Leggett_2001}, we call Rabi, Josephson and Fock.
These three regimes are characterized by a MZ phase sensitivity scaling as 
$N^{-1/2}$, $N^{-3/4}$ and $N^{-1}$, respectively, where $N$ is the number of particles in a single
interferometric experiment.
However, in a realistic dynamical splitting process, the $1/N$ limit requires 
a very long adiabatic ramping of the potential well. Its achievement is therefore strongly limited by the 
finite lifetime of the condensates. On the other hand, a sub shot-noise scaling $1/N^{3/4}$ can be actually 
reached. We restrict our discussion to a two-mode analysis, using a combination of numerical and analytical tools.

\newpage

\section{Two mode approximation}
In this section, we review a two-mode analysis of two weakly interacting BECs.
The second quantization Hamiltonian of a system of dilute bosons is given by 
\be 
\hat{H}(t)= \int \ud z \, \hat{\Psi}^{\dag}(z) \,
\bigg(-\frac{\hbar^2}{2m}\frac{\partial^2}{\partial z^2}+V(z,t)
\bigg) \, \hat{\Psi}(z)+ \frac{g}{2}\int \ud z \,
\hat{\Psi}^{\dag}(z) \hat{\Psi}^{\dag}(z) \hat{\Psi}(z) \hat{\Psi}(z), 
\ee 
where $\hat{\Psi}(z)$ is the bosonic field
operator, and $V(z,t)$ is the time-dependent external symmetric double-well potential, and $g= \frac{4 \pi \hbar^2 a}{m}$ 
is the strength of the interparticle interaction, with 
$a$ being the $s$-wave scattering length. The two-mode ansatz reads
\be \label{ansatz}
\hat{\Psi}(z)=\psi_a(z)\, \hat{a} + \psi_b(z)\, \hat{b}, 
\ee
where $\psi_{a,b}(z)$ can be constructed as sum and difference of the 
first symmetric and antisymmetric solutions of the Gross-Pitaevskii equation 
of a double well trap, correspondently to the 
ground and the first excited states, respectively.
The operators $\hat{a}^{\dag}$ and $\hat{b}^{\dag}$ ($\hat{a}$, $\hat{b}$)
create (destroy) a particle in the modes $a$, $b$, respectively. 
In the two-mode approximation the Hamiltonian of the system becomes
\cite{Javanainen_1999, Anglin_2001} 
\be \label{HBH}
\hat{H}=\frac{E_c}{4}\Big(\hat{a}^{\dag}\hat{a}^{\dag}\hat{a}\hat{a}+
\hat{b}^{\dag}\hat{b}^{\dag}\hat{b}\hat{b}\Big)
-K(t)\Big(\hat{a}^{\dag}\hat{b}+\hat{b}^{\dag}\hat{a} \Big). 
\ee 
where 
\be 
E_c=2 g \int \ud z \,|\psi_{a}(z)|^4=2 g \int \ud z \, |\psi_{b}(z)|^4, 
\ee
\be 
K(t)=-\int \ud z 
\bigg[\frac{\hbar^2}{2m}\frac{\partial \psi_a^*(z)}{\partial
z}\frac{\partial \psi_b(z)}{\partial z} + \psi_a^*(z) V(z,t) \psi_b(z) \bigg]
\ee
are the ``one-site energy'' and the ``Josephson coupling energy'', respectively.
The operator $\hat{N}=\hat{n}_a+\hat{n}_b=\hat{a}^{\dag}\hat{a}+\hat{b}^{\dag}\hat{b}$
is the total number of particles and commutes with $\hat{H}$.
By ramping the potential wells, $K(t)$ decreases with the 
decreasing of the overlap between the wave functions. For a linear ramping, a WKB 
approximation \cite{Zapata_1998} gives an exponential decrease $K(t)=K(0)e^{-t/\tau}$, where the effective ramping time 
$\tau=\frac{\Delta t_{\mathrm{ramp}}}{d\sqrt{V_0-\mu}}$ depends on the real ramping time 
$\Delta t_{\mathrm{ramp}}$, on the final distance $d$ between the wells, the height of the potential barrier 
$V_0$, and the chemical potential $\mu$ . 
Such a time-dependent configuration has been realized by the MIT group \cite{Shin_2004}. 

We study Eq.(\ref{HBH}) in a 
phase-states representation \cite{Anglin_2001}. We write a general state
in the Hilbert space of the two-mode system as
\be \label{SB}
|\psi\rangle=\int_{-\pi}^{+\pi}\frac{\ud \phi}{2\pi} \Psi(\phi,t) \, |\phi
\rangle,
\ee
where $\phi$ is the relative phase between the two modes, and 
\be \label{BV}
|\phi \rangle=\sum_{n=-N/2}^{N/2} \frac{e^{i n \phi}}
{\sqrt{\big(\frac{N}{2}-n\big)!}\sqrt{\big(\frac{N}{2}+n\big)!}} 
|N/2-n \rangle |N/2+n \rangle
\ee
are un-normalized vectors of the overcomplete phase basis, written in the relative number of 
particles $n$. In this representation the action of any two-mode operator applied to
$|\psi\rangle$ can be represented in terms of differential operators acting 
on the associated phase amplitude $\Psi(\phi,t)$. The main consequence of the overcompleteness is
the non-standard inner product between phase vectors (\ref{BV}) 
$\langle \, \phi \, | \, \theta \, \rangle = \frac{2^N}{N!}
\cos^N\big(\frac{\phi-\theta}{2}\big)$, which
affects both the inner product between states (\ref{SB}) and the mean values of observable.
By applying the Hamiltonian (\ref{HBH}) on the states (\ref{SB}) we obtain 
\be \label{Htot}
\hat{H} \, |\psi\rangle = \int_{-\pi}^{+\pi}\frac{\ud \phi}{2\pi} \, 
\big[ H_{eff}(\phi,t) \, \psi(\phi,t) \big]\, e^{\frac{2 K(t)}{E_c} \cos \phi} |\phi
\rangle,
\ee
where the effective Hamiltonian is
\be \label{Heff}
H_{eff}(\phi,t) =\bigg[-\frac{E_c}{2}\frac{\partial^2}{\partial \phi^2}-K(t)N \,\cos \phi - \frac{K^2(t)}{E_c} 
\cos 2 \phi \bigg]. 
\ee
In the first part of this work, we solve numerically and, in some limit, analytically
the eigenvalue equation 
\be \label{heis}
H_{eff}(\phi) \psi_{gs}(\phi)= E_{gs} \, \psi_{gs}(\phi),
\ee 
for different values of $K$. This will provide the limit of an adiabatic splitting of the condensate.  
In the second part we study the dynamical equation
\be \label{sch}
i \hbar \frac{\partial \psi(\phi,t)}{\partial t}= H_{eff}(\phi,t) \psi(\phi,t)
\ee
for different ramping times of the interwell barrier.
In both cases we give predictions of the Mach-Zehnder interferometer phase sensitivity, as discussed in
the following section.

\section{Mach-Zehnder interferometer}

For a compact analysis of the Mach-Zehnder interferometer, we 
introduce the Hermitian operators \cite{Yurke_1986, Campos_1989}
\begin{eqnarray}
\hat{J}_x=\frac{1}{2}\big(\hat{a}^{\dag}\hat{b}+\hat{b}^{\dag}\hat{a} \big) \\
\hat{J}_y=\frac{1}{2i}\big(\hat{a}^{\dag}\hat{b}-\hat{b}^{\dag}\hat{a} \big) \\
\hat{J}_z=\frac{1}{2}\big(\hat{a}^{\dag}\hat{a}-\hat{b}^{\dag}\hat{b} \big). \\
\nonumber
\end{eqnarray}
These operators form a SU(2) Lie algebra $[\hat{J}_i, \hat{J}_j]=i\epsilon_{i,j,k} \hat{J}_k$ and commute
with the total number of particles $\hat{N}$. 
The action of the Mach-Zehnder interferometer elements (beam splitter and phase shifter) on 
the vector $\hat{J}=(\hat{J}_x,\hat{J}_y,\hat{J}_z)$ can be represented by rotations in 
three-dimensional space. In particular, the whole interferometer can be represented by a 
rotation of $\hat{J}$ around the $y$-axis of an angle $\theta$ corresponding to the relative phase shift 
in the two arms. In our model, 
the information on the phase shift is inferred from the measure of the total $\hat{N}$ and the
relative mean $\langle \hat{J}_{z \, out} \rangle $ number of atoms in the two final condensates.
We have
\be \label{Jzout}
\langle \hat{J}_{z}^{\,out} \rangle=-\sin \theta \, \langle \hat{J}_x \rangle +
\cos \theta \, \langle \hat{J}_z \rangle
\ee
\be \label{DeltaJzout}
(\Delta\hat{J}_{z}^{\, out})^2=
\sin^2 \theta \, (\Delta\hat{J}_{x})^2+ \cos^2 \theta \, (\Delta\hat{J}_{z})^2 +
\sin \theta \cos \theta \, \big( 2 \langle \hat{J}_{x} \rangle \langle \hat{J}_{z} \rangle 
-\langle \hat{J}_{x} \hat{J}_{z} \rangle - \langle \hat{J}_{z} \hat{J}_{x} \rangle \big), 
\ee
where the expectation values are taken on the input state.
The sensitivity of the interferometer for the measure of the relative phase can be calculated by using the equation 
\cite{Yurke_1986}
\be \label{delta}
(\Delta \theta)^2=\frac{(\Delta \hat{J}_{z}^{\, out})^2}{|\partial \langle \hat{J}_{z}^{\,out} \rangle / \partial \theta|^2}.
\ee
For the optimal working pointy $\theta \sim 0$ and for perfect symmetric splitting of the condensate,
$\langle \hat{J}_z \rangle =0$. By using Eqs. (\ref{Jzout}, \ref{DeltaJzout}), we have 
\be \label{deltatheta}
(\Delta \theta)^2=\frac{\langle J_z^2 \rangle}{\langle J_x \rangle^2}.
\ee
This equation allows us to predict the sensitivity of the MZ interferometer
by knowing the input states. Because of its simplicity, this equation has been widely used in the literature 
to give predictions on the phase sensitivity of the 
Mach-Zehnder interferometer (see, for example \cite{Dowling_1998} and ref.s therein). 
We note, however, that this equation is based on the assumption that phase distributions are
Gaussian, a property that, in general, is not satisfied for states which are not coherent.
According to the central limit theorem, to obtain a Gaussian phase distribution, we have to combine several
independent measurements $p$, each having $N=N_T/p$ particles.
Therefore, equation (\ref{delta}) usually
gives the right scaling but a wrong prefactor. The exact MZ phase sensitivity can be obtained only with a rigorous
Bayesian analysis \cite{Helstrom}, which is rather cumbersome, and it will not be attempted here. 
We also remark that Eq.(\ref{delta}) is obtained in the linear case, i.e. with non-interacting particles.
This limit can be reached by switching off the interatomic interactions of the condensate after the creation of the
initial state. In the general case, Eq.(\ref{delta}) gives a lower bound for the MZ phase sensitivity. 
In the phase basis, we have:
\be
\hat{J}_x \, |\psi\rangle = \int_{-\pi}^{+\pi}\frac{\ud \phi}{2\pi} 
\, \Big[\sin \phi \frac{\partial}{\partial \phi}+\Big(\frac{N}{2}+1\Big) \cos \phi \Big] \, \Psi(\phi,t) \, |\phi \rangle,
\ee
\be
\hat{J}_y \, |\psi\rangle = \int_{-\pi}^{+\pi}\frac{\ud \phi}{2\pi} 
\, \Big[\cos \phi \frac{\partial}{\partial \phi}-\Big(\frac{N}{2}+1\Big) \sin \phi \Big] \, \Psi(\phi,t) \, |\phi \rangle,
\ee
\be
\hat{J}_z \, |\psi\rangle = \int_{-\pi}^{+\pi}\frac{\ud \phi}{2\pi} 
\, i \frac{\partial}{\partial \phi} \, \Psi(\phi,t) \, |\phi \rangle,
\ee
and
\be
\hat{J}^2_z \, |\psi\rangle = -\int_{-\pi}^{+\pi}\frac{\ud \phi}{2\pi} 
\, \frac{\partial^2}{\partial \phi^2} \, \Psi(\phi,t) \, |\phi \rangle.
\ee
We calculate the expectation values as in equation (\ref{deltatheta}), taking into account 
the overcomplete properties of this base. We find
\be \label{deltatheta2}
(\Delta \theta)^2
=\frac{-\int_{-\pi}^{+\pi}\frac{\ud \chi}{2\pi}\int_{-\pi}^{+\pi}\frac{\ud \phi}{2\pi} \, \langle \chi | \phi \rangle
\, \Psi^*(\chi,t) \big( \frac{\partial^2}{\partial \phi^2}  \Psi(\phi,t) \big)}
{\Big( \int_{-\pi}^{+\pi}\frac{\ud \chi}{2\pi}\int_{-\pi}^{+\pi}\frac{\ud \phi}{2\pi} \, \langle \chi | \phi \rangle
\, \Psi^*(\chi,t) \big[\sin \phi \frac{\partial}{\partial \phi}+(\frac{N}{2}+1) \cos \phi \big]\, \Psi(\phi,t)  \Big)^2  } 
\ee
where
\be \label{innerproduct}
\langle \chi | \phi \rangle = \frac{\cos^N\Big(\frac{\chi-\phi}{2} \Big)}
{\int_{-\pi}^{+\pi}\frac{\ud \chi}{2\pi}\int_{-\pi}^{+\pi}\frac{\ud \phi}{2\pi} 
\, \Psi^*(\chi,t) \, \Psi(\phi,t) \,\cos^N\Big(\frac{\chi-\phi}{2} \Big)  }.
\ee
Analytical calculations of the quantity (\ref{deltatheta2}) can be obtained in certain interesting limits. 
As we will discuss later, in the Rabi and Josephson regimes we can
consider a Gaussian phase amplitude 
$\Psi(\phi,t) \approx e^{-\phi^2/(4 \sigma^2)}$ and approximate the inner product as 
$\cos^N\big((\chi-\phi)/2\big)\approx e^{-N/2(\chi-\phi)^2}$.
If $\sigma \ll \pi$, we can extend the integral to $\pm \infty$, obtaining
\be \label{deltathetaapp}
(\Delta \theta)^2=\frac{\frac{N}{(1+4 N \sigma^2)}}{\Big[\big(\frac{N}{2}+1\big)
\big(1-\sigma^2\big(\frac{1+2 N \sigma^2}{1+4 N \sigma^2}\big)\big)-\big(\frac{1+2 N \sigma^2}{1+4 N \sigma^2}\big)
\Big]^2}.
\ee
In the opposite limit, when the phase uncertainty is of order of $2\pi$, we can neglect the inner product 
and consider the phase amplitude $\Psi(\phi,t) \approx (1+\gamma \cos \phi)$. In this limit, $\gamma \to 0$, we have 
\be \label{deltacos}
(\Delta \theta)^2=\frac{2+\gamma^2}{(N+2)^2}.
\ee
The whole phase sensitivity range can be exploited with the variational phase amplitude $\Psi(\phi,t) \approx e^{\gamma \cos \phi}$. We obtain 
\be \label{deltaexp}
(\Delta \theta)^2=\frac{2\gamma \int_{-\pi}^{+\pi} \ud \phi \, (\cos\phi)^N \big[\cos\phi \, \BI_1(2\gamma\cos\phi)
- \gamma (\sin\phi)^2 \, \big(\BI_0(2\gamma\cos\phi)+\BI_2(2\gamma\cos\phi)\big) \big]
\int_{-\pi}^{+\pi} \ud \phi \, (\cos\phi)^N \BI_0(2\gamma\cos\phi)}
{\Big(\int_{-\pi}^{+\pi} \ud \phi \,
(\cos\phi)^N \big[(N+1)\cos\phi \, \BI_1(2\gamma\cos\phi)- \gamma (\sin\phi)^2
\big(\BI_0(2\gamma\cos\phi)+\BI_2(2\gamma\cos\phi)\big)\Big)^2},
\ee
where $\BI_{\nu}(x)$ are Bessel functions of first kind and degree $\nu$.

\section{Adiabatic splitting}

Following the notation introduced in \cite{Leggett_2001} in the context of quantum tunneling 
between BECs in a two-well system, we can distinguish  
three main regimes depending on the ratio $K/E_C$ and the number of particles $N$.
As we will see, these three regimes
correspond to different input states of the MZ interferometer and, in particular, to three different 
scalings of the MZ phase sensitivity with the total number of particles. 

\begin{figure}[!ht]
\begin{center}
\includegraphics[scale=0.5]{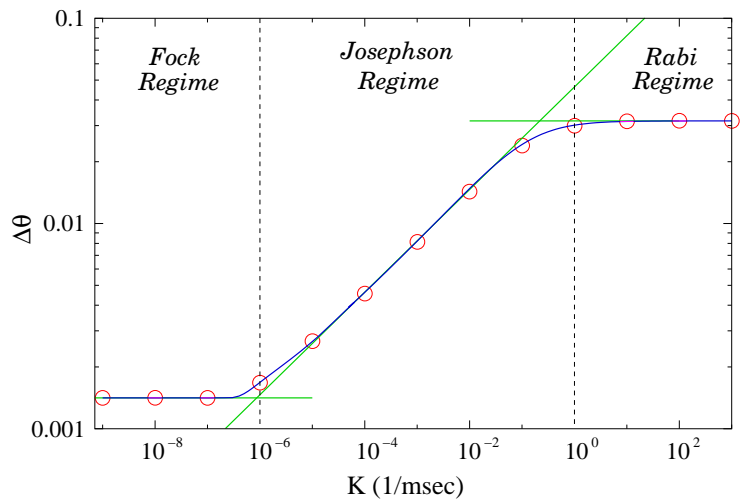}
\end{center}
\caption{\small{Color online. Mach Zehnder phase sensitivity $\Delta \Theta$
as a function of $K$ as given by Eq.(\ref{deltatheta2}). 
Here $N=1000$ and $E_c=0.001$ $msec^{-1}$. The green lines 
are the analytical predictions in three regimes: 
i) \emph{Rabi Regime}, $K\gg N Ec $, where 
$\Delta \theta=1/\sqrt{N}$, ii) \emph{Josephson Regime}, $Ec/N \ll K \ll N Ec$, where 
$\Delta \theta=(4 K/E_C)^{1/4} 1/N^{3/4}$, and iii) \emph{Fock Regime}, 
$K \ll Ec/N$, where $\Delta \theta=\sqrt{2}/N$. 
The blue line is given by the variational approach with the wave function (\ref{phaseamp}),
which reduced to Eq. (\ref{phasegaus}) in the Rabi and Josephson regimes.
Red points represent numerical solutions of Eq. (\ref{heis}).}}\label{adiabatic}
\end{figure}

\begin{description} 
\item[Rabi Regime]  $K/E_c \gg N $: This corresponds to a regime in which the two wells are not completely separated, 
the potential barrier is of the order of the chemical potential, and a strong tunneling exists
between the two condensates. 
In this limit, we can neglect the $\cos\phi$ potential term in Eq. (\ref{Heff}), and write
\be \label{Heff1}
H_{eff}(\phi) = -\frac{E_c}{2}\frac{\partial^2}{\partial \phi^2}- \frac{K^2}{E_c} \cos 2 \phi.
\ee
We note that this effective Hamiltonian does not depend on the total number of particles. 
A simple harmonic oscillator estimation gives
\be \label{sigma1}
\sigma^2_{\phi}=\frac{1}{4}\frac{E_c}{K},
\ee
which means that the phase dispersion is very small ($\sigma^2_{\phi}\ll 1/N$) and independent of $N$. The strong
tunneling characterizing this regime keeps the relative phase between the two condensates well defined.
However, the ground state of the 
Hamiltonian (\ref{Heff1}) is given by a function with peaks 
at $\phi=0$ and $\phi=\pm \pi$. To eliminate the unphysical peak at $\phi=\pm \pi$ it is necessary
to take into account the correction term $e^{\frac{2 K}{E_c} \cos \phi}$, 
contained in the full Hamiltonian Eq.(\ref{Htot}). 
This correction term has the same width (\ref{sigma1}) as the ground state of the Hamiltonian $H_{eff}$.
The corrected phase amplitude $\Psi(\phi)$ can be well approximated by a Gaussian of width given by half 
of Eq. (\ref{sigma1}):
\be \label{phasegaus}
\Psi(\phi)= \frac{e^{-\frac{\phi^2}{ 4 \frac{E_c}{8 k}}}}{\big(2 \pi \frac{E_c}{8 k}\big)^{1/4}}.
\ee
In the Rabi regime, taking into account that $N \sigma_{\phi}^2 \ll 1$, from equation (\ref{deltathetaapp}), we obtain 
\be
\Delta \theta=\frac{1}{\sqrt{N}},
\ee
which corresponds to the classical shot noise limit. 
We note that the phase amplitude (\ref{phasegaus}) does 
not depend on the total number of particles. The N dependence in $\Delta \theta$ is merely a 
consequence of 
the inner product (\ref{innerproduct}).
The sensitivity scaling as $\sim 1/\sqrt{N}$ is what we expect 
for a MZ interferometer fed by the coherent state 
$|\psi \rangle =(N!)^{-1/2}(\hat{a}^{\dag}+\hat{b}^{\dag})^N|\mathrm{vac}\rangle$.

\item[Josephson Regime] $\frac{1}{N} \ll \frac{K}{E_c} \ll N $: In this limit, we neglect the $\cos(2\phi)$ term
in the effective Hamiltonian  (\ref{Heff}), obtaining
\be 
H_{eff}(\phi) = -\frac{E_c}{2}\frac{\partial^2}{\partial \phi^2}- K N \cos \phi.
\ee
As in the previous regime, we can approximate the phase amplitude with a 
Gaussian of width 
\be \label{sigma2}
\sigma^2_{\phi}=\frac{1}{2}\sqrt{\frac{E_c}{K N}}.
\ee
In this case, however, the correction term $e^{\frac{2 K}{E_c} \cos \phi}$ gives a negligible contribution. 
From equation (\ref{deltathetaapp}), we obtain 
\be
\Delta \theta=\bigg(\frac{4 K}{E_c}\bigg)^{1/4}\frac{1}{N^{3/4}}.
\ee
We note that this result is recovered neglecting the inner product (\ref{innerproduct}). 
This is a consequence of the fact that in the Josephson regime the two functions $\psi_{gs}(\phi)$, and
$\tilde{\psi}_{gs}(\phi)=\int_{-\pi}^{+\pi} \ud \chi \, \psi_{gs}(\chi) \, \cos^N\big((\chi-\phi)/2\big)$ have the same 
width.

\item[Fock Regime] $\frac{K}{E_c} \ll \frac{1}{N}$: The effective Hamiltonian is still given by Eq. (\ref{Heff}), but 
now we have a nearly-free evolution. In this case the phase amplitude spreads over the whole $2 \pi$ interval and 
we can approximate it by 
\be \label{phaseamp}
\Psi(\phi)=\frac{e^{\gamma \cos \phi}}{\sqrt{2 \pi \BI_0(2 \gamma)}},
\ee
where $\gamma$ is a variational parameter and $\BI_0(2 \gamma)$ is the Bessel function of the first kind 
of degree zero. 
The behavior of $\gamma$ as a function of the parameters $E_c$, $K$, $N$, can be found by minimizing the
total energy
\be
E=\int_{-\pi}^{+\pi} \frac{\ud \phi}{2 \pi} \, \psi^*(\phi) 
\bigg[-\frac{E_c}{2}\frac{\partial ^2}{\partial \phi^2} - N K \cos\phi \bigg]  \psi^*(\phi), 
\ee
where we have not taken into account the inner product (\ref{innerproduct}). If we impose 
$\partial E/\partial \phi=0$, we obtain the equation
\be
\BI_0(2\gamma) \BI_1(2\gamma)+\bigg(\gamma-\frac{4 N K}{E_c}\bigg)\big[\BI_0^2(2\gamma)+\BI_0(\gamma)\BI_2(2\gamma)-2
\BI_1^2(2\gamma)\big]=0.
\ee
In the limit $\gamma \to 0$, we are in the Fock regime, the input state is given by 
$|\psi\rangle=(\hat{a}^{\dag})^{N/2}(\hat{b}^{\dag})^{N/2}|\mathrm{vac}\rangle$ and 
the phase amplitude becomes flat $\Psi(\phi)\to 1/\sqrt{2 \pi}$. 
The Fock state is characterized by a random phase \cite{Castin_1997}. This property renders the Fock 
state not useful in 
Young double-slit interferometry . In fact the interference fringes obtained in a single experiment
are washed out by statistical averaging over many experimental runs. 
In this limit, approximating Eq. (\ref{phaseamp}) with $\Psi(\phi)\approx 1+\gamma \cos \phi$, and by neglecting 
terms of order $o(\gamma^2)$ in (\ref{deltacos}), we obtain 
\be
\Delta \theta=\frac{\sqrt{2}}{N},
\ee
which corresponds to the Heisenberg limit of phase sensitivity.
Correcting to this equation to higher order in the variational parameter $\gamma$, it would be possible to 
exploit Eq. (\ref{deltaexp}), and express the Bessel functions $\BI_\nu(\gamma)$, $\nu=0,1,2$ with the series expansion
\be
I_\nu(\gamma)=\Big(\frac{\gamma}{2}\Big)^\nu \sum_{k=0}^{+ \infty} \frac{\big(\frac{\gamma}{2})^{2k}}{k!
\Gamma(k+1)}.
\ee
The corresponding MZ phase sensitivity is shown by the blue line of Fig.(\ref{adiabatic}).

\end{description}

In figure (\ref{adiabatic}) we plot the phase sensitivity with $N=1000$, $E_C=0.001\, msec^{-1}$, as a function of
$K$. We emphasize the three different regimes for the Mach Zehnder interferometry sensitivity characterized 
by a different scaling with the number of particles. We notice that the Josephson region becomes wider 
by increasing the number of particles $N$ and while keeping constant all other parameters. 
Therefore, strictly speaking, the phase sensitivity of the BEC MZ interferometer studied in this paper is bounded
by $1/N^{3/4}$ in the limit $N \to \infty$, when all other parameters are kept constant.

\section{Diabatic splitting}

\begin{figure}[!ht]
\begin{center}
\includegraphics[scale=0.78]{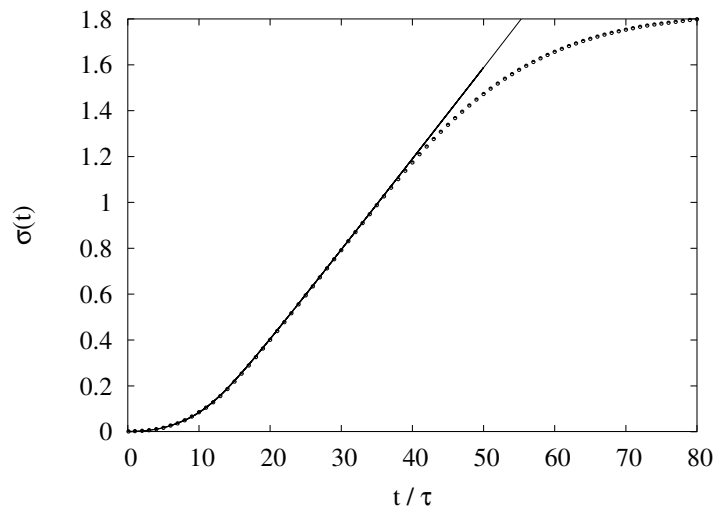}
\end{center}
\caption{\small{Comparison between the mean-square fluctuation $\sigma$, as obtained by the variational equations 
(\ref{sigmap},\ref{deltap})
(line) and the one obtained by the exact numerical solution of the Schr\"odinger equation (\ref{sch}) (points).
The parameters are $N=1000$, $E_C=0.001$ $msec^{-1}$, $K(0)=100$ $msec^{-1}$, and $\tau=10$ $msec$.}}\label{compsigma}
\end{figure}

We now study the dynamical splitting of the condensate by directly solving the Schr\"odinger equation (\ref{sch}), 
focusing on the dephasing process \cite{Pezze_2005}.
In particular, here we analyze
the sensitivity of the Mach-Zehnder interferometers fed by states created by a finite time, diabatic splitting.  
As shown in \cite{Pezze_2005}, initially the relative phase between the two split BECs 
has a very narrow distribution, corresponding to a BEC in a coherent state. 
While the height of the interwell barrier increases, $K(t)$ decreases, and the phase spreads in time over the whole $2\pi$ domain. 
Because of the periodic boundary conditions, the phase distributions eventually overlap around the region $\phi \sim \pm \pi$,
developing interference fringes with a wavelength increasing with $\tau$.
Using Eqs (\ref{deltatheta}, \ref{deltatheta2}), we study how the sensitivity of the MZ interferometer changes for 
different ramping times $\tau$.
The larger $\tau$, the more adiabatic the dynamics, and the closer it will follow 
the static model analyzed above (see Fig. (\ref{adiabatic})). 
If the breakdown of adiabaticity occurs in the Rabi-Josephson regime, the phase amplitude is sufficiently
narrow so
that we can study the dynamics with a variational approach using a Gaussian function \cite{Pezze_2005, Anglin_2001}.
We consider \cite{Smerzi_2000}
\be \label{variationalwf}
\psi(\phi,t)=\frac{1}{(2 \pi \sigma(t)^2)^{1/4}}e^{-\frac{\phi^2}{4
\sigma(t)^2}}e^{i\frac{\delta(t)}{2}\phi^2},
\ee 
where $\delta(t)$ and $\sigma(t)$ are time dependent variational parameters satisfying the differential equations
\be \label{sigmap}
\dot{\sigma}(t)=E_C \sigma(t) \delta(t) 
\ee
\be \label{deltap}
\dot{\delta}(t)=\frac{E_C}{4 \sigma^4(t)}-E_C \delta(t)^2-N\, k(t) e^{-\sigma^2/2}-\frac{4 k(t)}{E_C}e^{-2 \sigma^2}.
\ee
The variational wave function (\ref{variationalwf}) is in very good agreement with the 
numerical calculation in the regime $\sigma \leqslant 1$. In figure (\ref{compsigma}) we compare the mean-square fluctuation
$\sigma(t)$ obtained by the variational calculation (Eqs. (\ref{sigmap},\ref{deltap})), with the square root of the second moment 
of the phase amplitude obtained by the exact numerical solution of the Schr\"odinger equation (\ref{sch}). 
The agreement is very good until the phase amplitude touches the borders $\pm \pi$. As discussed in \cite{Pezze_2005}, this
happens when $\sigma \approx 1$.

\begin{figure}[!ht]
\begin{center}
\includegraphics[scale=0.6]{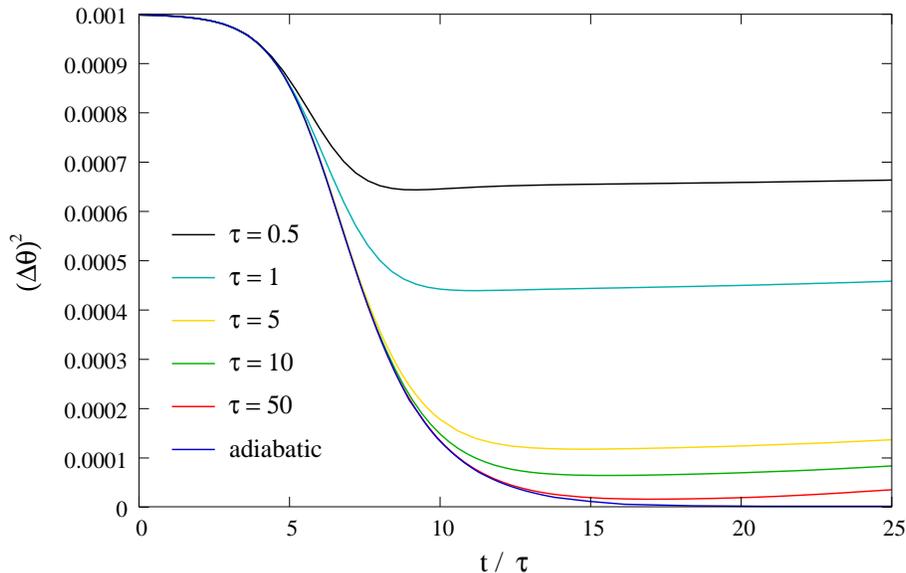}
\end{center}
\caption{\small{Color online. Plot of $(\Delta \theta)^2$ as given by the equation (\ref{deltatheta2}) for different
values of $t/\tau$ (in msec). The parameters are $N=1000$, $E_C=0.001$ $msec^{-1}$ and $K(0)=100$ $msec^{-1}$. 
The blue line represents the adiabatic behavior}}\label{Dphi}
\end{figure}

In a realistic experimental setup, the two condensates initially are in the Rabi regime. 
While increasing the interwell barrier, the phase amplitude adiabatically follows the 
decreasing of the effective potential energy in Eq. (\ref{Heff}), which is proportional to $K(t)^2$. 
After the breakdown of adiabaticity $t_{ad}$, the phase distribution continues to spread but at a slower rate. 
When the Josephson coupling energy is sufficiently small, the dynamical evolution becomes essentially free,  
and the phase uncertainty increases as a consequence of excitations of the system. 
In figure (\ref{Dphi}) we plot the MZ phase sensitivity $(\Delta \theta)^2$ (Eq. (\ref{deltatheta2})) for different values 
of $\tau$; the blue line represents the adiabatic behavior. The dynamics are calculated with the variational wave function
(\ref{variationalwf}). Notice that the minimum of the phase fluctuation is below the point of breakdown of adiabaticity.

\begin{figure}[!ht]
\begin{center}
\includegraphics[width=10 cm, height=7 cm]{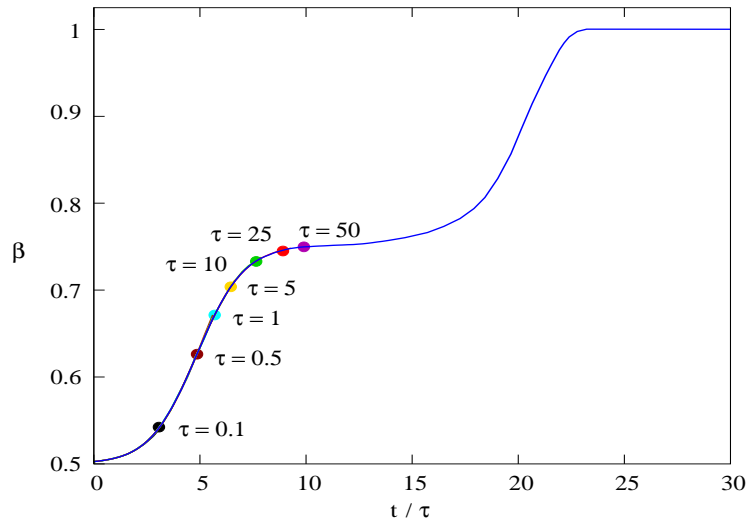}
\end{center}
\caption{\small{Color online. Plot of the scaling parameter $\beta$ with the number of particles (see eq.
(\ref{betad})). The calculation has been made with $N=10000$ and $N=1000$, and 
with fixed parameters $E_C=0.001$ $msec^{-1}$, and $K(0)=100$ $msec^{-1}$.
The blue line represents the adiabatic behavior; the points correspond to the minimum of MZ phase sensitivity
occurring for different values of $\tau$ (in $msec$).}}\label{beta}
\end{figure}

We now study the MZ sensitivity by parametrizing the phase error as
\be \label{betad}
(\Delta \theta)^2 = \frac{\alpha}{N^{\beta}}.
\ee 
Here we assume that the prefactor $\alpha$ does not depend of the number of particles $N$, as we have numerically
verified in the limit $N \gg 1$.
In fig. (\ref{beta}) we plot the quantity $\beta$ as a function of $t/\tau=\ln(K(0)/K)$ obtained 
by calculating the scaling between
the cases $N=10000$ and $N=1000$, with fixed parameters $E_C=0.001$ $msec^{-1}$, and $K(0)=100$ $msec^{-1}$.
The blue line represents the adiabatic behavior, and we can clearly distinguish
the three regions described above. The points in fig. (\ref{beta}) correspond to the MZ sensitivity at the breakdown  
of adiabaticity.  As we can see, the limit $1/N$ can be reached only with a very long adiabatic ramping, a
circumstance that is strongly limited by the finite lifetime of the condensate. However, the sub shot-noise limit $1/N^{3/4}$
can be achieved under currently available experimental conditions \cite{Shin_2004, Oberthaler_2005}.
We remark that, by increasing N while keeping the same initial conditions, there are two interesting effects that 
take place: i) the minimum of the phase uncertainty tends to coincide with the breakdown of adiabaticity, 
and ii) the increase of the phase uncertainty after reaching the minimum is rather slow. 
Both these two effects can be understood if we approximate the evolution of the phase amplitude after $t_{ad}$ 
with a free expansion model ($k(t)=0$ for $t>t_{ad}$) \cite{Leggett_1998}. In this case 
\be
\sigma^2(t)=\sigma^2(t_{ad})+\frac{E_C}{4 \hbar \sigma^2(t_{ad})}(t-t_{ad})^2.
\ee
The bigger $\sigma^2(t_{ad})$, the slower the dynamics of the phase amplitude.  
In the temporal range $t_{ad} \leq t \lesssim t_{ad}+2 \sigma^2(t_{ad}) \hbar/E_C$, the dynamical
spreading is almost frozen. On the other hand, if
we increase $N$ by keeping the other initial conditions $K(0)$, $E_C$ and $\tau$ constant, the breakdown of adiabaticity
occurs at larger values of $\sigma(t_{ad})$ \cite{Javanainen_1999}. As a consequence, by increasing N the
MZ sensitivity will freeze after the breakdown of adiabaticity, and the minimum of phase sensitivity is maintained for a longer
time.\\
 
\section{Conclusions}
We have studied the phase sensitivity of a Mach-Zehnder interferometer fed by quantum states 
produced by the splitting a single Bose Einstein Condensate in two parts. We studied the process in the two-mode
approximation, and we analyzed the sensitivity of the MZ interferometer by projecting into an overcomplete phase basis
and by using an error propagation formula. 
We have first calculated the Mach-Zehnder phase sensitivity in the adiabatic splitting limit.
We distinguished three different regimes
(Rabi, Josephson, and Fock) characterized by a different scaling of the phase sensitivity with 
the number of particles in a single experiment. While the $1/N$ scaling can hardly be reached in realistic
experiments, the limit $\sim 1/N^{3/4}$, corresponding to the Josephson regime, can be achieved with current 
technology \cite{Shin_2004, Oberthaler_2005}, offering a considerable improvement in phase sensitivity over the 
shot noise $1/\sqrt{N}$ obtainable in the classical limit. \\
This work  was supported by the Department of Energy under the
contract W-7405-ENG-36 and DOE Office of Basic Energy Sciences.

Pfister,O./Holland,M.J./Noh,J./Hall,J.L./

\end{document}